**Thickness-dependent quantum transport of Weyl fermions in ultra-high-quality SrRuO₃ films**

Shingo Kaneta-Takada,[1,2,*] Yuki K. Wakabayashi,[1,*,†] Yoshiharu Krockenberger,[1] Shinobu Ohya,[2,3] Masaaki Tanaka,[2,4] Yoshitaka Taniyasu,[1] and Hideki Yamamoto[1]

[1]*NTT Basic Research Laboratories, NTT Corporation, Atsugi, Kanagawa 243-0198, Japan*
[2]*Department of Electrical Engineering and Information Systems, The University of Tokyo, Bunkyo, Tokyo 113-8656, Japan*
[3]*Institute of Engineering Innovation, The University of Tokyo, Bunkyo, Tokyo 113-8656, Japan*
[4]*Center for Spintronics Research Network (CSRN), The University of Tokyo, Bunkyo, Tokyo 113-8656, Japan*

[*]These authors contributed equally to this work.
[†]Author to whom correspondence should be addressed: yuuki.wakabayashi.we@hco.ntt.co.jp

Abstract

The recent observation of Weyl fermions in the itinerant 4*d* ferromagnetic perovskite SrRuO₃ points to this material being a good platform for exploring novel physics related to a pair of Weyl nodes in epitaxial heterostructures. In this letter, we report the thickness-dependent magnetotransport properties of ultra-high-quality epitaxial SrRuO₃ films grown under optimized conditions on SrTiO₃ substrates. Signatures of Weyl fermion transport, i.e., unsaturated linear positive magnetoresistance accompanied by a quantum oscillation having a π Berry phase, were observed in films with thicknesses as small as 10 nm. Residual resistivity increased with decreasing film thickness, indicating disorder near the interface between SrRuO₃ and the SrTiO₃ substrate. Since this disorder affects the magnetic and electrical properties of the films, the Curie temperature decreases and the coercive field increases with decreasing thickness. Thickness-dependent magnetotransport measurements revealed that the threshold residual resistivity ratio (RRR) to observe Weyl fermion transport is 21. These results provide guidelines for realizing quantum transport of Weyl fermions in SrRuO₃ near heterointerfaces.

The itinerant ferromagnetic perovskite $SrRuO_3$ has been widely used in oxide electronics and spintronics as an epitaxial conducting layer owing to the unique nature of its ferromagnetism, metallicity, compatibility with other perovskite-structured oxides, and chemical stability.[1–17] The reasons for renewed interest in $SrRuO_3$ are twofold. One is the observation of quantum transport of Weyl fermions in $SrRuO_3$,[18] such as chiral-anomaly-induced negative magnetoresistance (MR), unsaturated linear positive MR, $\pi$ Berry phase along cyclotron orbits, light cyclotron masses, and high quantum mobility of about 10000 $cm^2$/Vs. The other is the realization of two-dimensional ferromagnetism in electronically conducting $SrRuO_3$ thin films of one unit-cell thickness, embedded in $[(SrRuO_3)_1/(SrTiO_3)_n]$ heterostructures.[19,20] These attractive phenomena have been observed only in samples of exceptional quality.[18,19] The residual resistivity ratio (RRR) is a good measure to gauge the purity of metallic systems, that is, the quality of single-crystalline $SrRuO_3$ thin films.[8,11,18] It is known that proper Ru stoichiometry control during growth is necessary to obtain $SrRuO_3$ thin films with a high RRR,[19] since RRR rapidly decreases with increasing Ru vacancies.[8,11] A high RRR value is essential to revealing the intrinsic electronic structure and exploiting the full potential of $SrRuO_3$.[18,19] However, only a few papers have reported $SrRuO_3$ films with RRRs over 50 due to difficulties in fine-tuning multiple growth conditions.[7,10,14,18] Thus, a thorough understanding of its transport properties, electronic structure, and origin of its ferromagnetism has remained elusive despite tremendous efforts for over the past five decades.[8]

For future investigations into $SrRuO_3$-based heterostructures aimed at applying magnetic Weyl semimetals to spintronic devices[21] and topoelectrical circuits,[22,23] it will be essential to realize Weyl transport near heterointerfaces. In epitaxial $SrRuO_3$ films, higher resistivity and lower Curie temperature ($T_C$) are commonly found when the film thickness is decreased.[8,24–28] Therefore, there should be a critical thickness for observing the transport signatures of Weyl fermions, depending on the quality of $SrRuO_3$ films.

In this letter, we present the magnetotransport properties of epitaxial $SrRuO_3$ films with various thicknesses (1–60 nm), grown on $SrTiO_3$ substrates. The values of RRR above 60 for thick films are a hallmark of ultra-high-quality films, enabling observation of quantum transport of Weyl fermions.[18] Signatures of quantum transport of Weyl fermions, i.e., an unsaturated linear positive MR accompanied by a quantum oscillations having a $\pi$ Berry phase, were observed in $SrRuO_3$ films with thicknesses of 10 nm or more. We also found that the threshold RRR to observe Weyl fermion transport in thickness-dependent magnetotransport is 20.7, which coincides well with the value (19.4) in crystallinity-dependent magnetotransport for 63-nm-thick films.[18] These results serve as guidelines for accessing quantum transport of Weyl fermions in $SrRuO_3$-based heterostructures.

We grew high-quality epitaxial $SrRuO_3$ films with various thicknesses $t$ (= 1–60 nm) on (001) $SrTiO_3$ substrates in a custom-designed molecular beam epitaxy (MBE) setup equipped with multiple e-beam evaporators for Sr and Ru. The growth parameters were optimized by Bayesian optimization, a machine learning technique for parameter optimization,[29–31] with which we achieved RRR > 60 when the film thickness is large (60 nm). We precisely controlled the elemental fluxes, even for elements with high melting points, e.g., Ru (2250°C), by monitoring the flux rates with an electron-impact-emission-spectroscopy sensor, which was fed back to the power supplies for the e-beam evaporators. The Ru and Sr fluxes were 0.365 and 0.980 Å/s, respectively, corresponding to Ru-rich conditions. Excessive Ru is known to be reevaporated from the growth surface through formation of volatile species such as $RuO_4$ and $RuO_3$ under oxidizing atmosphere,[11] leading to stoichiometric films. The growth rate of 1.05 Å/s was deduced from the thickness calibration of a thick (63 nm) $SrRuO_3$ film using cross-sectional scanning transmission electron microscopy (STEM). This growth rate agrees very well with the value (1.08 Å/s) estimated from the flux rate of Sr, confirming the accuracy of the film thickness. The oxidation during growth was carried out with a mixture of ozone ($O_3$) and $O_2$ gas (~15% $O_3$ + 85% $O_2$), which was introduced at a flow rate ~2 sccm through an alumina nozzle that was pointed at the substrate. All $SrRuO_3$ films were grown at 772°C. Further information about the MBE setup and preparation of the substrates are described elsewhere.[32−34] Magnetotransport was measured by using a standard four-probe method with Ag electrodes deposited on the $SrRuO_3$ surface without any additional processing. The distance between the two voltage electrodes was 2 mm.

The surface morphology of the 63-nm-thick $SrRuO_3$ film with an RRR of 51.8 consisted of atomically flat terraces and steps, as observed by atomic force microscopy in our previous study.[31] The out-of-plane lattice constant estimated from the Nelson-Riley extrapolation method for $\theta$-$2\theta$ XRD patterns of the film with $t = 60$ nm was 3.949 Å, which is ~0.5% larger than the pseudocubic lattice constant of bulk specimens (3.93 Å).[8] This implies that the $SrRuO_3$ film was compressively strained due to the lattice constant of the $SrTiO_3$ substrate (3.905 Å) being ~0.6% smaller than the pseudocubic lattice constant of $SrRuO_3$.

Figure 1(a) shows RHEED patterns of $SrRuO_3$ surfaces for various thicknesses $t$. These films show sharp streaky patterns, which indicate the growth of the $SrRuO_3$ film proceeded in a two-dimensional layer-by-layer manner leading to high crystalline quality. Since rough surface morphology due to three-dimensional growth enhances electron scattering, a two-dimensional growth mode is necessary to achieve a large electronic conduction for ultra-thin $SrRuO_3$ films below 2 nm.[19,24,25,28] Figure 1(b) shows a cross-sectional high-angle annular dark-field scanning transmission electron microscopy (HAADF-STEM) image of the film with $t = 63$ nm as an example. $SrRuO_3$ grew epitaxially on a (001) $SrTiO_3$ substrate with an abrupt substrate/film interface, as expected from the RHEED patterns.

The temperature ($T$) dependence of the longitudinal resistivity $\rho_{xx}$ of the $SrRuO_3$ films measured using a standard four-probe method is shown in Fig. 2(a). The $\rho_{xx}$ of the films with $t = 5$–60 nm decreased with decreasing temperature, indicating that these films were metallic in the whole temperature range. Similarly, the $\rho_{xx}$ of the $t = 1$ and 2 nm films decreased with decreasing temperature from 300 K. Below the temperature, $T_{\rho min}$, at which $\rho_{xx}$ reached a minimum, $\rho_{xx}$ increased with decreasing temperature, indicating insulating behavior plausibly due to weak localization in the low-temperature region.[19,24−28] Although the thickness-dependent insulating behavior of $SrRuO_3$ films has been reported by many other groups,[19,24−28] in most of those studies, insulating behavior starts to appear when the $SrRuO_3$ thickness is 2–4 nm.[24−27] Meanwhile, in our study, the film with $t = 2$ nm was metallic, indicating that fine-tuning of the growth conditions is essential to obtaining metallic $SrRuO_3$ films thinner than 2 nm through the suppression of disorder due to Ru vacancies and other defects.[31] The inset in Fig. 2(b) plots $\rho_{xx}$ versus $T^2$ and shows a linear fitting for $t = 60$ nm as an example. The films with $t = 5$–60 nm showed a $T^2$ scattering rate ($\rho_{xx} \propto T^2$) that is expected for a Fermi liquid, in which electron-electron scattering dominates the transport;[8,14,18] specifically, transport phenomena intrinsic to a Fermi liquid appeared below this temperature (hereafter called $T_F$) except for temperature-independent residual resistivity arising from elastic scattering due to, e.g., disorder. Figure 2(b) shows the $t$ dependence of the residual resistivity $\rho_{Res}$ at 2 K and $T_F$. $\rho_{Res}$ increased and $T_F$ decreased with decreasing $t$, and the Fermi liquid behavior disappeared at $t < 2$ nm [Fig. 2(b)]. As mentioned above, clear insulating behavior appeared at $t = 1$ nm [see the inset in Fig. 2(a)]. These results—the increase in $\rho_{Res}$ with decreasing $t$ as well as the low-temperature upturn in the thinnest film—suggest that there is disorder near the interface between $SrRuO_3$ and $SrTiO_3$ substrate, which is rather insensitive to cross-sectional STEM measurements. Figure 2(c) plots the $t$ dependence of RRR [$= \rho_{xx}(300\ \mathrm{K})/\rho_{xx}(2\ \mathrm{K})$]. RRR monotonically decreased from 60.1 to 2.4 with decreasing $t$ from 60 to 2 nm, reflecting the increase in $\rho_{Res}$ near the interface.

When $t \geq 5$ nm, the $\rho_{xx}$ vs. $T$ curves show clear kinks [arrows in Fig. 2(a)], at which the ferromagnetic transition occurs and spin-dependent scattering is suppressed.[8] To highlight this transition, we plot the derivative resistivity $d\rho_{xx}/dT$ as a function of $T$ in Fig. 2(d). Here, we define $T_C$ as the temperature at which the $d\rho_{xx}/dT$ curves show a clear peak or kink. Note that, in general, $T_C$ determined from $d\rho_{xx}/dT$ is a few K lower than the values measured from the temperature dependence of the magnetization.[26] As shown in Figs. 2(e), $T_C$ gradually decreases with decreasing $t$ ($\geq 10$ nm), with an abrupt change below 10 nm. Since $\rho_{Res}$ is very sensitive to Ru vacancies and other defects,[8,11] the low $T_C$ values in the thin samples can be attributed to these sources of disorder near the interface, which should degrade the exchange interaction in $SrRuO_3$.

We performed magnetotransport measurements on the $SrRuO_3$ films by using the standard four-probe method. Figure 3(a) shows the $t$ dependence of MR (($\rho_{xx}(B)-\rho_{xx}(0\ \mathrm{T}))/\rho_{xx}(0\ \mathrm{T})$) when the magnetic field $B$ was applied in the out-of-plane [001] direction of the $SrTiO_3$ substrate at 2 K. The sign of MR at high magnetic fields changed from negative to positive with increasing $t$ [Fig. 3(a)]. Importantly, the linear positive MR at 2 K showed no signature of saturation even up to 9 T, which is commonly seen in Weyl semimetals[35–40] and is thought to stem from the linear energy dispersion of Weyl nodes.[41,42] Furthermore, films with $t \geq 10$ nm showed quantum oscillations in resistivity [i.e., Shubnikov-de Haas (SdH) oscillations], whose frequency (26 T) corresponds to that of Weyl fermions in $SrRuO_3$, as described later [see also the inset in Fig 3(a)].[18] In our previous study,[18] we confirmed the existence of Weyl fermions in $SrRuO_3$ by observing five signatures of Weyl fermions in quantum transport:[36] a linear positive MR, chiral-anomaly-induced negative MR, $\pi$ phase shift in quantum oscillations, light cyclotron mass, and high quantum mobility of about 10000 cm²/Vs, and by performing first-principles electronic structure calculations. Thus, in $SrRuO_3$, the unsaturated linear positive MR and SdH oscillations, which appear simultaneously when $t \geq 10$ nm, are clear signatures of Weyl fermion transport. From the results shown in Fig. 3(a), we extracted the $t$ dependence of the MR ratio at 2 K with 9 T applied in the out-of-plane [001] direction of the $SrTiO_3$ substrate [Fig. 3(b)]. The MR ratio increases with increasing $t$, and a linear positive MR is observed when $t \geq 10$ nm. We also plotted the RRR dependence of the MR ratio [Fig. 3(c)]; the MR ratio increased with increasing RRR. This means that Weyl fermions become more dominant in the transport when $SrRuO_3$ has fewer defects and higher crystalline quality, since it is easily hindered by disorder.[18] Unsaturated linear positive MR was observed in films with RRR $\geq 20.7$. This threshold value of RRR for the appearance of Weyl fermion transport is consistent with our previous study on 63-nm-thick $SrRuO_3$ films, in which unsaturated linear positive MR was observed in films with RRR = 19.4.[18]

In Fig 3(a), anisotropic magnetoresistance (AMR) hysteresis is also observed,[43] which is proportional to the relative angle between the electric current and the magnetization, occurred below 3 T at 2 K, irrespective of film thickness. In SrRuO₃, the position of the AMR peak corresponds to the coercive field $H_c$.[18] Figure 3(d) shows that $H_c$, as estimated from the positions of the AMR peaks in Fig. 3(a), increases with decreasing thickness below 20 nm. Since the magnetic domains tend to be pinned by grain boundaries and other defects, the large $H_c$ likely stems from the poor crystalline quality near the interface.

Finally, we analyzed the SdH oscillations for Weyl fermions in the SrRuO₃ films. Figures 4(a) and 4(b) show the analysis of the SdH oscillation in the film with $t$ = 10 nm. We subtracted the background by using a polynomial function up to fifth order from Fig. 4(a) and extracted the oscillation components, as shown in Fig. 4(b). For the Fourier transformation, we interpolated the data of Fig. 4(b) to prepare equally spaced $x$-axis ($1/B$) points. Then, we multiplied the data with a Hanning window function to obtain the periodicity of the above data. Finally, we conducted a fast Fourier transform on the data [Fig. 4(c)]. SdH oscillations with a frequency of ~26 T, which corresponds to Weyl fermions in SrRuO₃,[18] were observed in the films with $t \geq 10$ nm. A $\pi$ Berry phase along the cyclotron orbits was also detected through the SdH oscillation.[18,36,38–40] Note that, over the past three decades, the Berry phase has become an important concept in condensed matter physics, since it represents the topological classification of the system. [44] According to Lifshitz-Kosevich theory, a non-zero Berry phase appears in the phase shift of the SdH oscillation. The magnitude of the SdH oscillation is described as[45]

$$\Delta\sigma_{xx} = A \frac{X}{\sinh X} \exp\left(-\frac{2\pi^2 k_B T_D}{\hbar\omega_c}\right) \cos\left[2\pi\left(\frac{F}{B} - \frac{1}{2} + \beta_B\right)\right],$$

where $\Delta\sigma_{xx}$ is the oscillation component of the longitudinal conductivity, $A$ is the normalization factor, $X = 2\pi^2 k_B T/\hbar\omega_c$, $k_B$ is the Boltzmann constant, $\hbar$ is the reduced Planck constant, $\omega_c$ is the cyclotron frequency defined as $eB/m^*$, $m^*$ is the cyclotron mass, $T_D$ is the Dingle temperature, $F$ is the frequency of the SdH oscillation, and $2\pi\beta_B$ is the phase shift caused by the Berry phase. To extract the Berry phases in SrRuO₃, we used the above formula to fit to the SdH oscillation data. In the fitting, we fixed $T_D$ and $m^*$ to 0.63 K and $0.35m_0$ ($m_0$: electron rest mass), i.e., the values obtained in our previous study.[18] As shown in Fig. 4(b), the signal of the SdH oscillation is well reproduced by using $\beta_B$ =0.59. The SdH oscillation cannot be reproduced by a zero Berry phase, which confirms the existence of a non-zero Berry phase [Fig. 4(b)]. The $2\pi\beta_B$ value of $1.18\pi$ (=2π×0.59) comes from the nontrivial character of the massless dispersion of Weyl fermions.[18,36,38–40] As described above, the SdH oscillations of the Weyl fermions and the unsaturated linear positive MR appear simultaneously when $t \geq$ 10 nm. These results emphasize that the critical thickness for observing quantum transport of Weyl fermions in our SrRuO₃ films is as small as 10 nm.

In summary, we have investigated the thickness dependence of the magnetotransport properties in ultra-high-quality SrRuO₃ films ($t$ = 1–60 nm). The sign of MR changed from negative to positive with increasing thickness, and the signature of Weyl fermion transport, i.e. unsaturated linear positive MR accompanied by SdH oscillations having a non-trivial Berry phase, was observed at $t \geq$ 10 nm. The threshold RRR to observe Weyl fermion transport in thickness-dependent magnetotransport was 20.7, in good agreement with the value (19.4) deduced from 63-nm-thick films with various crystallinities. Realizing a Weyl fermion transport in ultra-thin $SrRuO_3$ films with $t < 10$ nm may require further improvement in crystallinity near the interface. For example, superlattices of $[(SrRuO_3)_1/(SrTiO_3)_n]$ appear to be promising means of refining crystallinity at $SrRuO_3/SrTiO_3$ interfaces.[19] Our results present important insights for applying Weyl fermions in $SrRuO_3$-based heterostructures to spintronic devices and topoelectrical circuits. The ability to make high-quality $SrRuO_3$ films will open the door to topological oxide electronics.

## ACKNOWLEDGEMENTS

We thank Kosuke Takiguchi for valuable discussions. Part of this work was conducted at the Cryogenic Research Center of the University of Tokyo.

## AUTHORS' CONTRIBUTIONS

Y.K.W. conceived the idea, designed the experiments, and led the project. Y.K.W. and Y.K. grew the samples. S.K.T. and Y.K.W. carried out the sample characterizations. S.K.T. and Y.K.W. carried out the transport measurements and analyzed the data. All authors contributed to the discussion of the data. S.K.T. and Y.K.W. co-wrote the paper with input from all authors.

## DATA AVAILABILITYY

The data that support the findings of this study are available from the corresponding author upon reasonable request.

**Figures and figure captions**

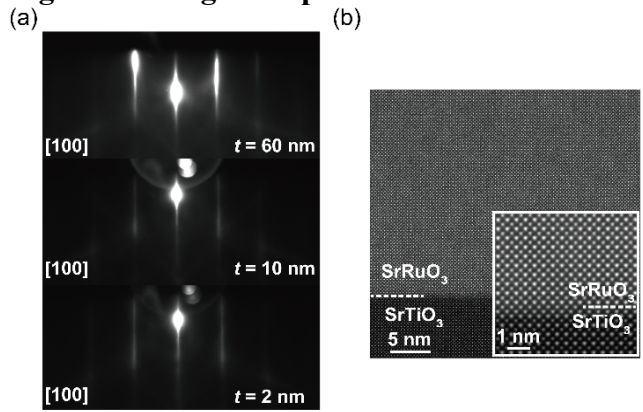

FIG. 1. (a) RHEED patterns of SrRuO₃ films with $t$ = 60, 10 and 2 nm taken along the [100] axis of the SrTiO₃ substrates. (b) HAADF-STEM images of SrRuO₃ films taken along the [100] axis of the SrTiO₃ substrates. The inset in (b) is a magnified image near the interface.

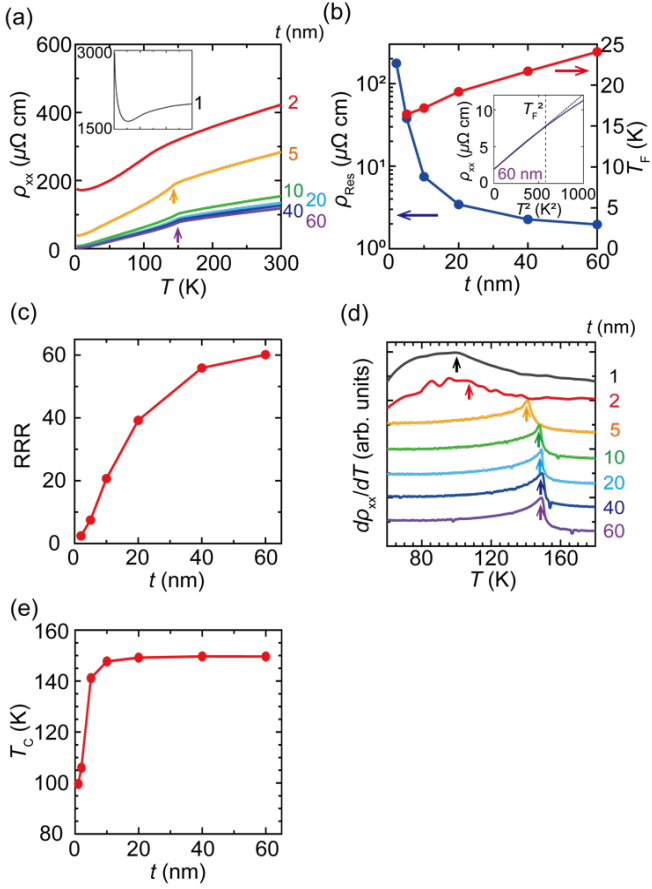

FIG. 2. (a) Temperature dependence of resistivity $\rho_{xx}$ on thickness $t$ (=1–60 nm). Arrows indicate kinks. The inset in (a) shows data for $t = 1$ nm. (b) Thickness $t$ dependence of residual resistivity $\rho_{Res}$ and $T_F$. The inset in (b) is a $\rho_{xx}$ versus $T^2$ plot with a linear fitting (black dashed line) for $t = 60$ nm. (c) Thickness $t$ dependence of RRR. (d) Temperature dependence of differential resistivity $d\rho_{xx}/dT$ for various $t$ (=1–60 nm). Arrows indicate the positions of the peaks and kinks in the $d\rho_{xx}/dT$ curves. In (d), $d\rho_{xx}/dT$ has been offset for easy viewing. (e) Thickness $t$ dependence of $T_C$ obtained from the positions of peaks in differential resistivity $d\rho_{xx}/dT$ in (d).

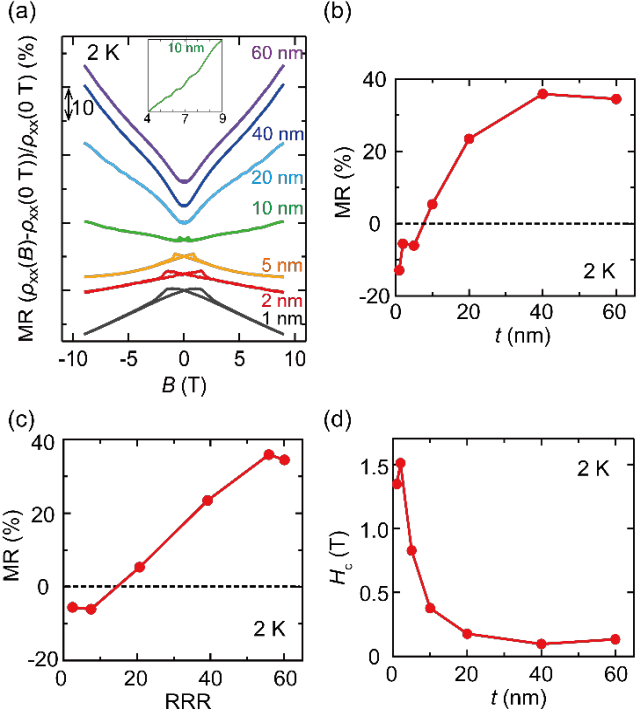

FIG. 3. (a) Thickness $t$ dependence of MR $(\rho_{xx}(B)-\rho_{xx}(0\ T))/\rho_{xx}(0\ T)$ at 2 K with $B$ applied in the out-of-plane [001] direction of the SrTiO$_3$ substrate. The inset in (a) shows the SdH oscillation with $t$ = 10 nm with $B$ (4 T < $B$ < 9 T). (b),(c) Thickness $t$ (b) and RRR (c) dependence of the MR ratio at 2 K with $B$ = 9 T applied in the out-of-plane [001] direction of the SrTiO$_3$ substrate. (d) Thickness $t$ dependence of coercive field $H_c$ obtained from the positions of the MR peaks in (a).

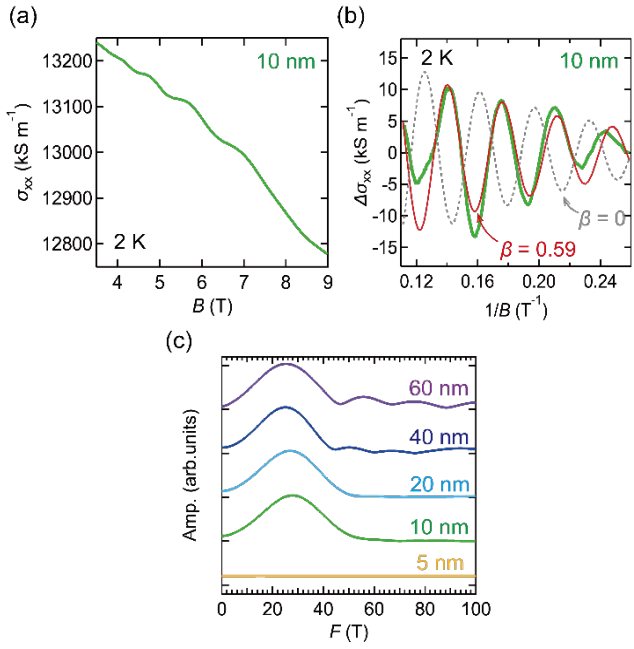

FIG. 4. (a) SdH oscillation in raw $\rho_{xx}$ data at 2 K with $B$ (3.5 T < $B$ < 9 T) applied in the out-of-plane [001] direction of the SrTiO₃ substrate for SrRuO₃ film with $t$ = 10 nm. (b) Background-subtracted SdH oscillation at 2 K with $B$ (4 T < $B$ < 9 T) for film with $t$ = 10 nm. (c) Fourier transform spectra of SdH oscillation at 2 K for films with $t$ = 5–60 nm.